\documentclass{jfm}

\pdfoutput=1

\usepackage{url}
\usepackage{color}
\usepackage{soul}
\usepackage[english]{babel}
\usepackage{amssymb}
\usepackage{amsmath}
\usepackage{bm}
\usepackage{subfigure}
\usepackage{natbib}
\usepackage{graphicx}
\usepackage{blkarray}
\usepackage{array}
\usepackage{color}
\usepackage[utf8]{inputenc}
\usepackage[T1]{fontenc}

\begin{document}

\shorttitle{Subcritical turbulent condensate in RRRBC}
\shortauthor{Favier et al.}

\title{Subcritical turbulent condensate in rapidly rotating Rayleigh-B\'enard convection}

\author
 {
 Benjamin Favier\aff{1}
  \corresp{\email{favier@irphe.univ-mrs.fr}},
 C\'eline Guervilly\aff{2}
  and
 Edgar Knobloch\aff{3}
  }

\affiliation
{
\aff{1}
Aix Marseille Univ, CNRS, Centrale Marseille, IRPHE, Marseille, France
\aff{2}
School of Mathematics, Statistics and Physics, Newcastle University, UK
\aff{3}
Department of Physics, University of California, Berkeley, CA 94720, USA
}

\maketitle

\begin{abstract}
The possibility of subcritical behaviour in the geostrophic turbulence regime of rapidly rotating thermally driven convection is explored. In this regime a non-local inverse energy transfer may compete with the more traditional and local direct cascade. We show that, even for control parameters for which no inverse cascade has previously been observed, a subcritical transition towards a large-scale vortex state can occur when the system is initialized with a vortex dipole of finite amplitude. This new example of bistability in a turbulent flow, which may not be specific to rotating convection, opens up new avenues for studying energy transfer in strongly anisotropic three-dimensional flows.
\end{abstract}

\section{Introduction}

Turbulence in geophysical and astrophysical systems is a problem of major importance.
Three-dimensional (3D) flows favour a forward cascade, i.e., energy flows from large to small scales, as described by the well-known Kolmogorov theory. In contrast, two-dimensional (2D) flows exhibit an inverse energy transfer, from small to large scales, that manifests itself in the appearance of large scale structures in the flow \citep{kraichnan1967,boffetta2012}. However, many systems arising in geophysical and astrophysical fluid dynamics fail to be fully 3D because of the presence of strong restraints, although they are far from being 2D either. These restraints may arise from geometrical confinement \citep{smith1996,celani2010,benavides_alexakis_2017,xia2017}, rapid rotation \citep{smith1999,pouquet2013b,campagne2014}, strong stratification \citep{bart1995,smith2002,oks2017} or the presence of strong magnetic fields \citep{Alexakis2011,favier_godeferd_cambon_delache_bos_2011}. The detailed nature of the energy cascade in constrained 3D flows and its applications to geophysical fluid dynamics remain an open problem \citep{ALEXAKIS2018}.

Recent numerical simulations of rapidly rotating Rayleigh-B\'enard convection, hereafter RRRBC, have shown that the turbulent state is susceptible to the evolution of large scale vortex (LSV) structures despite the presence of 3D fluctuations on all scales \citep{guervilly_hughes_jones_2014,favier2014pof}, in agreement with the predictions of an asymptotic description of the system valid in the limit of vanishingly small Rossby numbers \citep{julien2012,rubio2014}. Together these studies reveal the presence of an efficient energy transfer mechanism that extracts energy from small scale 3D fluctuations and deposits it in a box-scale barotropic (i.e. 2D) mode, bypassing the inverse energy cascade familiar from 2D dynamics. This process operates in parallel to energy transfer to large scale 2D modes by barotropic-barotropic interactions \citep{rubio2014} but dominates all aspects of the problem. \cite{julien2012} and \cite{julien2018} have suggested that the presence of the LSV introduces essential correlations among the phases of the small scale 3D fluctuations that facilitate direct energy extraction from these scales by the large scale mode, leading to a runaway that is only arrested by additional processes omitted from the simplest problem formulation. Such a runaway is characteristic of subcritical dynamics where it is triggered by finite amplitudes perturbations. The present paper is therefore devoted to a search for such subcritical behaviour in RRRBC. We do not address the question of whether geostrophic turbulence is itself linearly unstable to the generation of an LSV structure although our simulations suggest that at large enough rotation rates and large enough Rayleigh numbers this is in fact the case. We emphasize that phase correlations are missed in studies that focus on energy spectra alone, and that such correlations are inevitably absent from flows driven by a prescribed small scale force such as those studied by \cite{chertkov} and \cite{bouchet}. In these systems the only possible correlations are between the applied force and the resulting velocity field. In contrast, in the present 3D system the forcing of the large scale 2D flow can itself be dynamically affected through the action of the LSV on the small convective scales. The LSV observed in RRRBC may thus be a consequence of the proximity of the flow to 2D turbulence or due to the ability of the LSV to shape the correlations among the small scale fluctuations that appear to drive its formation. 

In the present work we provide the first evidence for subcritical dynamics in turbulent RRRBC by demonstrating the coexistence of two numerically stable turbulent states at identical parameter values, one with an LSV structure and one without. Such bistability in turbulent flows is rare although it has also been found in rapidly rotating turbulence \citep{alexakis_2015,Yokoyama2017}, thin-layer turbulence \citep{van2018condensates}, Couette flows \citep{mujica,zimmerman2011,lohse,xia_shi_cai_wan_chen_2018} and von K\'arm\'an flows \citep{Ravelet2004}.

\vspace{-4mm}
\section{Mathematical formulation}

\subsection{Model and governing equations}

We consider the evolution of a layer of incompressible fluid, bounded above and below by two impenetrable, fixed temperature, stress-free horizontal walls, a distance $h$ apart. The layer rotates about the $z$-axis, pointing vertically upwards, with a constant angular velocity $\bm{\Omega}=\Omega\bm{e}_z$ while gravity points downwards: $\bm{g}=-g\bm{e}_z$.
The kinematic viscosity $\nu$ and thermal diffusivity $\kappa$ are assumed to be constant.

In the Boussinesq approximation, using the thermal diffusion time $h^2/\kappa$ as a unit of time and the depth $h$ of the layer as a unit of length, the dimensionless equations are
\begin{equation}
\label{eq:momentum}
\frac{1}{Pr}\left(\frac{\partial\bm{u}}{\partial t}+\bm{u}\cdot\nabla\bm{u}\right)=-\nabla p-\sqrt{Ta}\,\bm{e}_z\times\bm{u}+Ra \theta\, \bm{e}_z+\nabla^2\bm{u} \ ,
\end{equation}
\vspace{-6mm}
\begin{equation}
\label{eq:div}
\nabla\cdot\bm{u}=0 \ ,
\end{equation}
\vspace{-6mm}
\begin{equation}
\label{eq:temp}
\frac{\partial\theta}{\partial t}+\bm{u}\cdot\nabla\theta=w+\nabla^2\theta \ ,
\end{equation}
where $\bm{u}\equiv \left(u, v, w\right)$ is the velocity, $p$ is the pressure and $\theta$ is the temperature fluctuation with respect to a linearly decreasing background temperature.
The parameters are the Rayleigh number $Ra=\alpha g \Delta Th^3/(\nu\kappa)$, the Taylor number $Ta=4\Omega^2h^4/\nu^2$ and the Prandtl number $Pr=\nu/\kappa$.
These dimensionless quantities involve $\alpha$, the coefficient of thermal expansion, and $\Delta T$, the imposed temperature difference between the two horizontal plates. 
For simplicity, and following earlier studies of the formation of large scales structures in this system, we take $Pr=1$.
In the two horizontal directions, all variables are assumed to be periodic with the same spatial period $\lambda$ in both $x$ and $y$ directions. The boundary conditions at the upper ($z=1$) and lower ($z=0$) plates are $\partial_zu = \partial_z v = w = \theta = 0$.

We solve equations~\eqref{eq:momentum}-\eqref{eq:temp} using the same mixed Fourier fourth-order finite difference scheme as used in \cite{favier2014pof}. We confirm the robustness of the present results by additionally running equivalent simulations using the fully spectral approach of \cite{guervilly_hughes_jones_2014} and the open-source spectral-element code Nek5000 developed by \cite{nek5000} at the Argonne National Laboratory.

\setlength{\tabcolsep}{5.5pt}
\begin{table*}
\centering
\caption{Summary of the parameters considered: $Ta$ is the Taylor number, $Ra$ is the Rayleigh number, $Ro=\sqrt{Ra/(Pr Ta)}$ is the input Rossby number, $\widetilde{Ra}=Ra Ta^{-2/3}$ is a scaled Rayleigh number and $\lambda$ is the horizontal aspect ratio. The Prandtl number $Pr=1$ in all simulations.\label{tab:one}}
\begin{tabular}{@{}c|cccccccc@{}}
\hline
\hline
Case & $Ta$ & $Ra$ & $Ro$ & $\widetilde{Ra}$ & $\lambda$ & $N_x \! \times \! N_y \! \times \! N_z$ & $A$ & Subcriticality\\
\hline
A1 & $10^8$ & $3\times10^7$ & $0.55$ & $139$ & $2$ & $256^3$ & $[0:1000]$ & Yes\\
A2 & $10^8$ & $3\times10^7$ & $0.55$ & $139$ & $4$ & $512^2\times 256$ & $[0:1600]$ & Yes\\
A3 & $10^8$ & $3\times10^7$ & $0.55$ & $139$ & $6$ & $768^2\times 256$ & $800, 1200$ & Yes \\[0.0cm]
B1 & $10^8$ & $4\times10^7$ & $0.63$ & $186$ & $2$ & $256^3$ & $[1000:4000]$ & No\\
B2 & $10^8$ & $4\times10^7$ & $0.63$ & $186$ & $4$ & $512^2\times 256$ & $2000$ & Yes\\[0.0cm]
C1 & $10^8$ & $5\times10^7$ & $0.71$ & $232$ & $2$ & $256^3$ & $[1000:4000]$ & No\\[0.0cm]
D1 & $10^8$ & $4\times10^6$ & $0.2$ & $18.6$ & $4$ & $256^2\times128$ & $[0:1000]$ & No\\[0.0cm]
\hline
\hline
\end{tabular}
\end{table*}

\subsection{Finite amplitude initial conditions}

In order to explore possible subcritical behaviour in turbulent RRRBC, we consider a particular set of initial conditions. Anticipating that the non-local upscale energy transfer eventually saturates by creating a barotropic vortex dipole at the box scale \citep{julien2012,rubio2014}, we consider a depth-invariant initial condition, given by
\begin{equation}
\label{eq:init}
    \bm{u}=\Big(A\sin\left(2\pi y/\lambda\right),-A\sin\left(2\pi x/\lambda\right),0\Big) \quad \quad \text{and} \quad \quad \theta=0 \ ,
\end{equation}
and parametrised by the amplitude $A$.
This initial condition corresponds to a symmetric vortex dipole at the box scale with a cyclone located at the box center $x=\lambda/2$ and $y=\lambda/2$ and an anticyclone located at the corners of the periodic domain.
In addition, we add random perturbations of small amplitude $\pm2.5\times10^{-2}$ to the temperature field in order to initiate the Rayleigh-B\'enard instability.
The total kinetic energy density of this initial flow is
\begin{equation}
    K_0=\frac1V\int_V\frac12\bm{u}\cdot\bm{u} \ \text{d}V=\frac{1}{2}A^2\,,
\end{equation}
where $V$ is the total volume.
Note that $K_0$ is independent of the aspect ratio of the box.
The purely viscous decay of this initial condition is given by
\begin{equation}
\label{eq:viscous_decay}
    K(t)=K_0\exp(-8\pi^2 Pr \ t/\lambda^2) \ .
\end{equation}

\subsection{Flow decomposition}

We define the depth-averaged 2D horizontal flow, subsequently called the 2D mode, as
\begin{equation}
\label{eq:vortex}
\left<u\right>_z\!(x,y) = \int_0^1 u(x,y,z) \, \textrm{d}z \quad \quad \text{and} \quad \quad \left<v\right>_z\!(x,y) = \int_0^1 v(x,y,z) \, \textrm{d}z \ ,
\end{equation}
where $u$ and $v$ are the velocity components in the $x$ and $y$ directions, respectively, and fast 3D fluctuations, subsequently called the 3D mode, as
\begin{align}
\label{eq:upr}
u'(x,y,z) & = u(x,y,z)-\left<u\right>_z(x,y)\,, \\
\label{eq:vpr}
v'(x,y,z) & = v(x,y,z)-\left<v\right>_z(x,y)\,.
\end{align}
We can then define the (purely horizontal) kinetic energy density associated with the slow 2D mode as
\begin{equation}
\label{eq:k2d}
K_{2D}=\frac{1}{2\lambda^2}\iint\left(\left<u\right>_z^2+\left<v\right>_z^2\right) \ \textrm{d}x \ \textrm{d}y
\end{equation}
and the kinetic energy density associated with the fast 3D mode as
\begin{equation}
\label{eq:k3d}
K_{3D}=\frac{1}{2\lambda^2}\iiint\left(u'^2+v'^2+w^2\right) \ \textrm{d}x \ \textrm{d}y \ \textrm{d}z \ .
\end{equation}

\begin{figure}
   \vspace{2mm}
   \centering
   \hspace{-4mm}
   \includegraphics[width=0.5\textwidth]{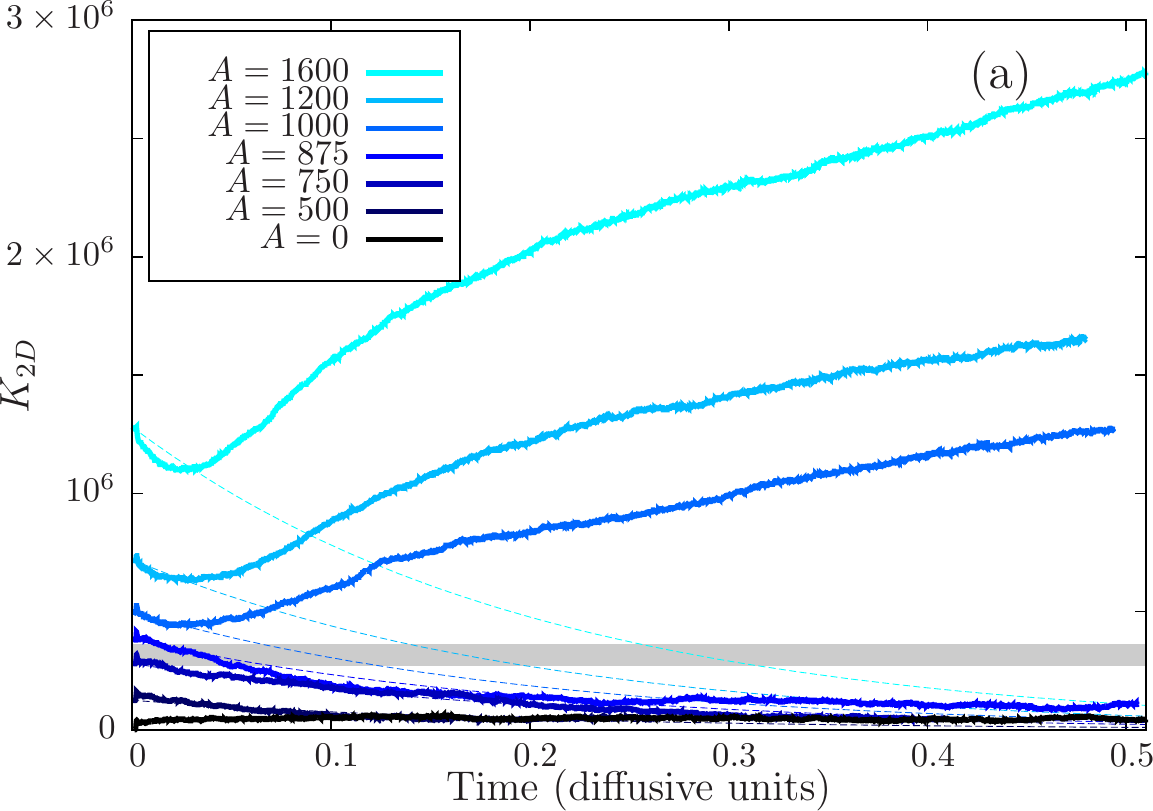}
   \hspace{0.0mm}
   \includegraphics[width=0.505\textwidth]{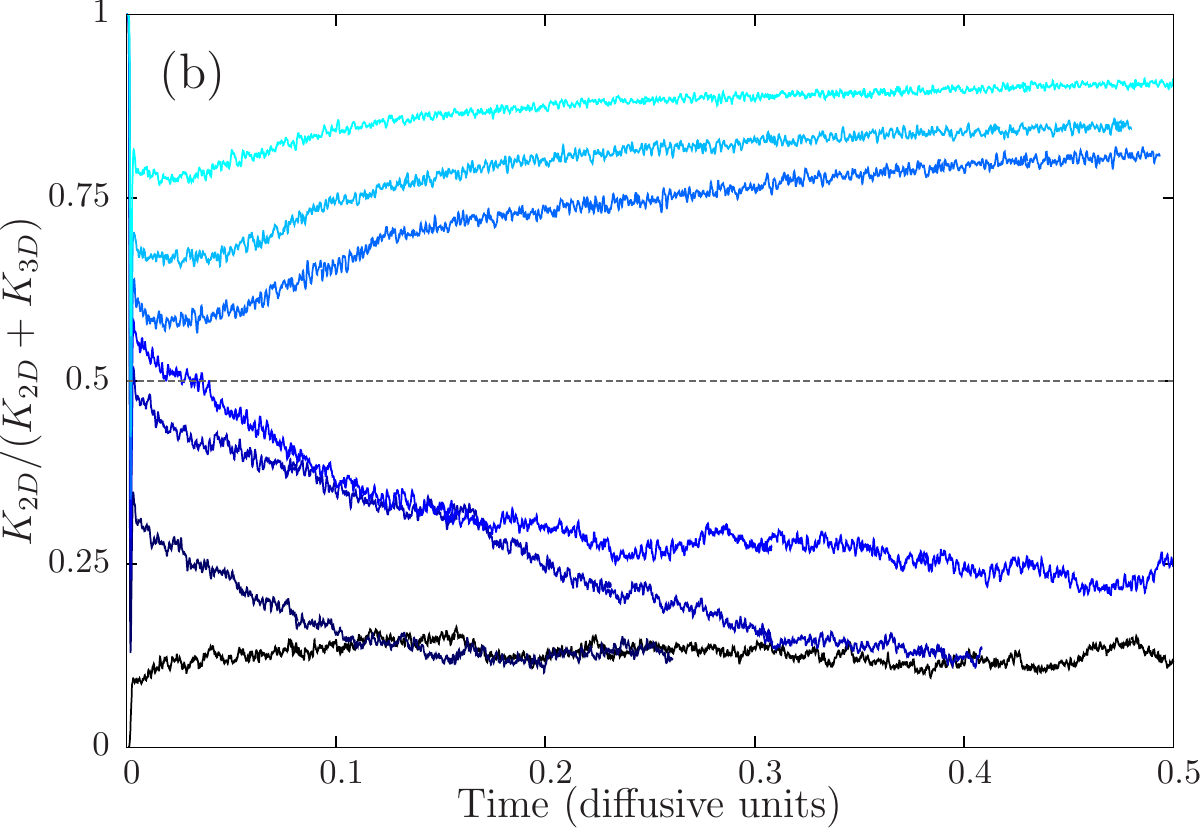}
   \caption{(a) Time evolution of the kinetic energy density $K_{2D}$ for different initial vortex amplitudes $A$. (b) Time evolution of the ratio between $K_{2D}$ and the total kinetic energy $K_{2D}+K_{3D}$ for different amplitudes $A$. The grey area corresponds to the transition where $K_{2D}\approx K_{3D}$.\label{fig:kin1}}
\end{figure} 

\section{Results}

\begin{figure}
   \vspace{2mm}
   \centering
   \includegraphics[width=0.38\textwidth]{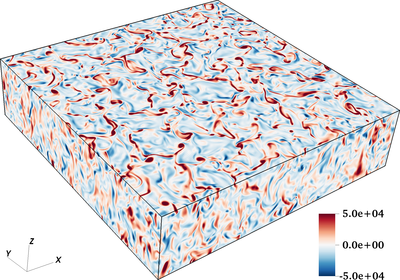}
   \hspace{1.5cm}
   \includegraphics[width=0.38\textwidth]{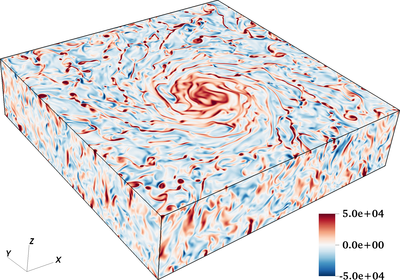}\\
   \vspace{-0mm}
   \includegraphics[width=0.38\textwidth]{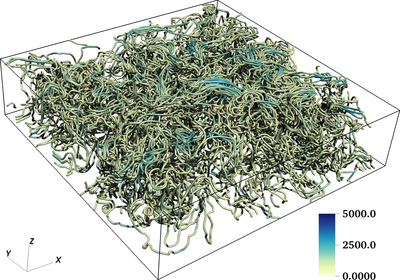}
   \hspace{1.5cm}
   \includegraphics[width=0.38\textwidth]{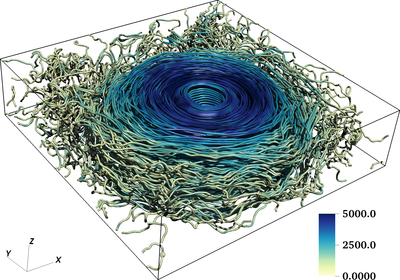}
   \caption{Visualizations of the quasi-steady states at $t=0.3$. Top: vertical vorticity component. Bottom: velocity field streamlines colored with the velocity amplitude. Left: no initial vortex dipole $A=0$. Right: initial vortex dipole amplitude $A=1600$.\label{fig:visus}}
\end{figure}

In this paper, we fix $Ta=10^8$ which is sufficient to sustain a non-local inverse cascade for $5\times10^6\lesssim Ra\lesssim2\times10^7$ \citep{guervilly_hughes_jones_2014,favier2014pof}. For $Ra<5\times10^6$ the flow is not turbulent enough while for $Ra>2\times10^7$ the flow is insufficiently constrained by rotation and so not anisotropic enough, and only the traditional forward energy cascade is observed.
In between these two limits, the flow is both turbulent and dynamically constrained by rotation, leading to the spontaneous emergence of a LSV from purely 3D perturbations driven by the Rayleigh-B\'enard instability.
In order to better understand the nature of the transition from a state with LSV to a state without, we consider here the particular case $Ra=3\times10^7$, for which no systematic LSV were observed starting from random infinitesimal temperature perturbations \citep{guervilly_hughes_jones_2014,favier2014pof}, and focus on the behaviour with aspect ratio $\lambda=4$.
This aspect ratio is chosen to ensure a clear scale separation between the convective eddies and the box size (see section \ref{sec:discuss} for a discussion of the effects of varying $\lambda$). For this set of control parameters, the flow is dominated by 3D small scale turbulent fluctuations (see left panels in figure~\ref{fig:visus}), characterized by the Reynolds number $Re=\sqrt{\left<w^2\right>}\approx535$ and the micro-Rossby number $Ro_{\omega}=\sqrt{\left<\omega_z^2\right>}/\sqrt{Ta}\approx1.4$. Here $\omega_z$ is the vertical component of the vorticity and the brackets denote spatial and temporal averaging.
Using this simulation as a reference, we ran a number of additional simulations, continuously varying the amplitude $A$ of the initial vortex dipole defined in Eq.~\eqref{eq:init}.

Figure~\ref{fig:kin1}(a) shows the temporal evolution of the kinetic energy density $K_{2D}$ of the 2D flow as defined by Eq.~\eqref{eq:k2d}.
We observe a clear transition as the initial amplitude $A$ of the vortex dipole increases.
For small amplitudes, typically $A\lesssim800$, the kinetic energy of the 2D flow decreases until it reaches the equilibrium value corresponding to the reference case $A=0$ (i.e., no initial vortex).
Note that this decay closely follows the purely viscous decay of the initial condition as given by Eq.~\eqref{eq:viscous_decay}, shown as dashed lines in figure~\ref{fig:kin1}(a), indicating that there is no significant energy transfer from the 3D fluctuations to the 2D flow.
For larger amplitudes however, typically $A\gtrsim800$, we observe an initial decay of $K_{2D}$ followed by an approximately linear increase until the energy eventually saturates at very long times.
Note that close to the transition threshold, see case $A=875$ for example, it is not yet clear whether the vortex will grow or decay.
In view of figure \ref{fig:kin1}(a) which shows that the large scale vortex has not yet reached saturation, additional and longer simulations are required to determine any residual dependence of the saturated state on $A$ although we expect that all solutions involving a growing LSV will eventually saturate at the same amplitude.

These results clearly point towards subcritical behaviour in the transition between a weak subdominant 2D flow and a strong non-local inverse energy transfer efficiently feeding energy into the largest available spatial scale of the domain. Note that while the two states correspond to the same control parameters, the LSV state has a total kinetic energy density approximately 8 times that of the 3D state.
The ratio between $K_{2D}$ and the total kinetic energy $K_{2D}+K_{3D}$ is shown in figure~\ref{fig:kin1}(b).
It appears that the finite amplitude transition occurs once that $K_{2D}(t=0)\approx K_{3D}$ although it depends very subtly on the initial condition and can occur even after a long transient, as in the case $A=875$.
A more accurate estimate of the critical amplitude $A$ is beyond the scope of this paper, however, since it would require running many realisations only changing the initial temperature perturbation (see section~\ref{sec:discuss} below for a brief discussion concerning the nondeterministic nature of this transition).
Visualizations of the saturated states for the same control parameters, but two different initial conditions, are shown in figure~\ref{fig:visus}.
Without the initial LSV, or when the amplitude $A$ is too small, the equilibrium state is dominated by a 3D flow at small scales while the 2D flow remains marginal.
These small scale fluctuations are fully 3D, as expected from the moderate value of the Rossby number, $Ro=0.55$.
In contrast, above the critical value of $A$, the LSV is continuously amplified while remaining at the box scale.

\begin{figure}
   \vspace{2mm}
   \centering
   \includegraphics[width=0.49\textwidth]{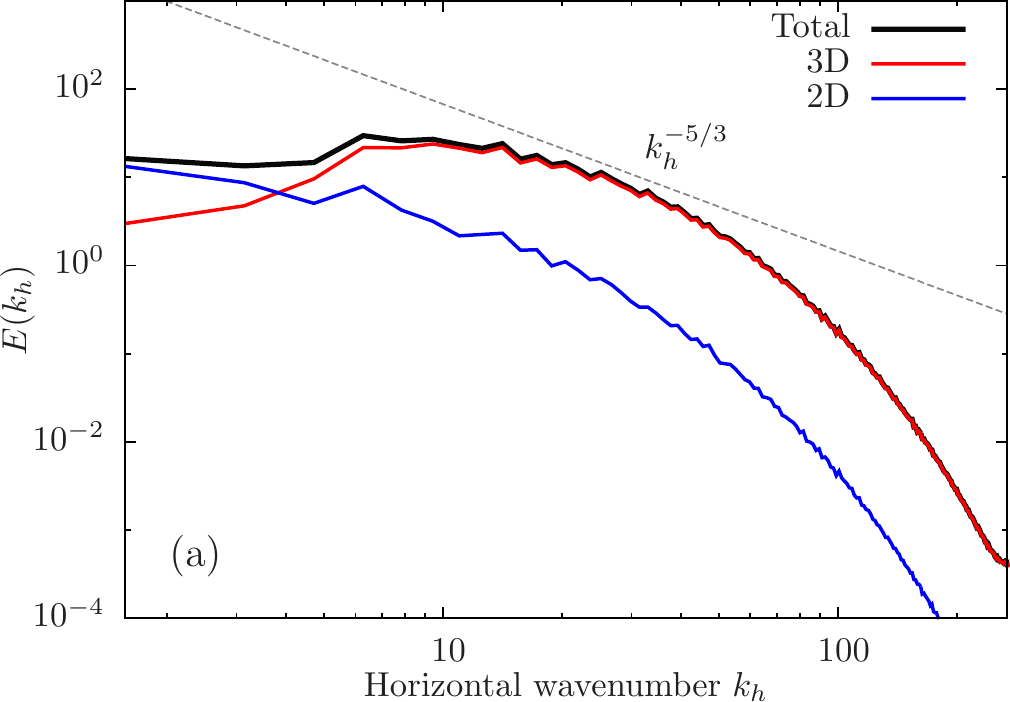}
   \hfill
   \includegraphics[width=0.49\textwidth]{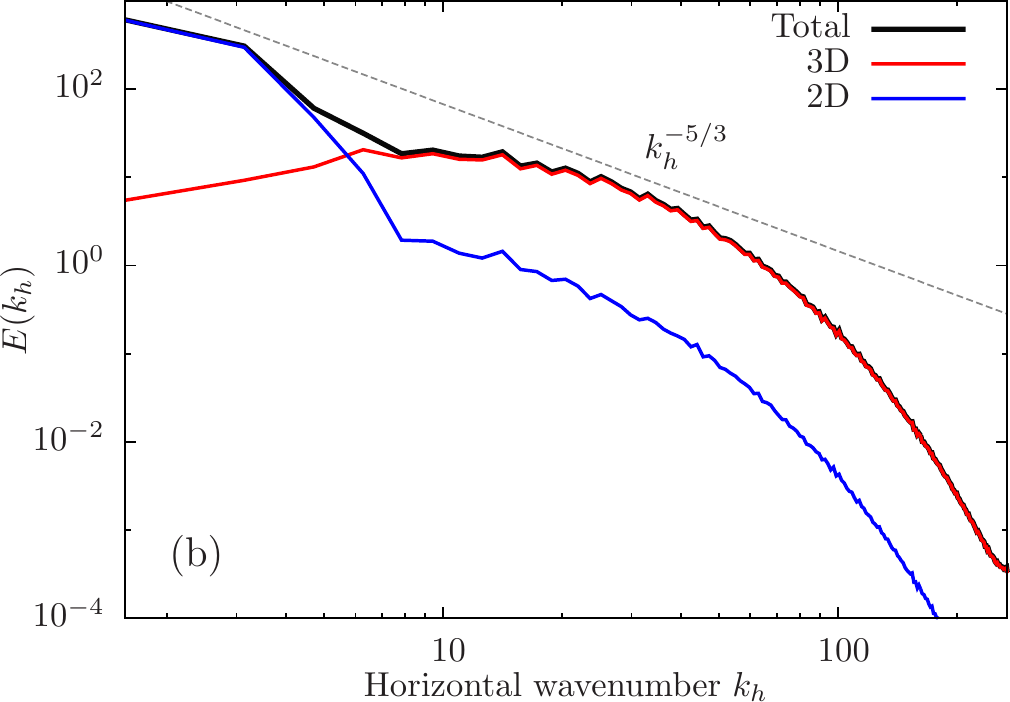}
   \caption{Kinetic energy spectra averaged over depth ($0<z<1$) and time ($0.3<t<0.4$) as a function of the horizontal wavenumber $k_h\equiv\sqrt{k_x^2+k_y^2}$. (a) $A=0$. (b) $A=1600$.\label{fig:spect1}}
\end{figure}

We now focus on the spectral statistics of the two different states.
The kinetic energy spectra for each component of the flow, as defined by equations~\eqref{eq:vortex}-\eqref{eq:vpr} (see also \cite{guervilly_hughes_jones_2014} and \cite{favier2014pof}), are shown in figure~\ref{fig:spect1}.
These spectra are averaged over depth and time.
With no initial vortex, the 3D flow is dominant at virtually all scales except for the smallest available wavenumber where most of the energy is contained in the 2D mode.
This subdominant 2D flow is in equilibrium, meaning that there is a balance between viscous dissipation and baroclinic forcing (see below).
There is no systematic growth of the 2D mode and no large scale condensate is reached.
Above the critical initial amplitude, however, the LSV is able to extract energy efficiently from the small scale 3D flow leading to rapid growth of the 2D mode with horizontal wavenumber $k_h\leq3$, while the 3D flow remains largely unchanged at all scales. 

\begin{figure}
   \vspace{2mm}
   \centering
   \includegraphics[width=0.49\textwidth]{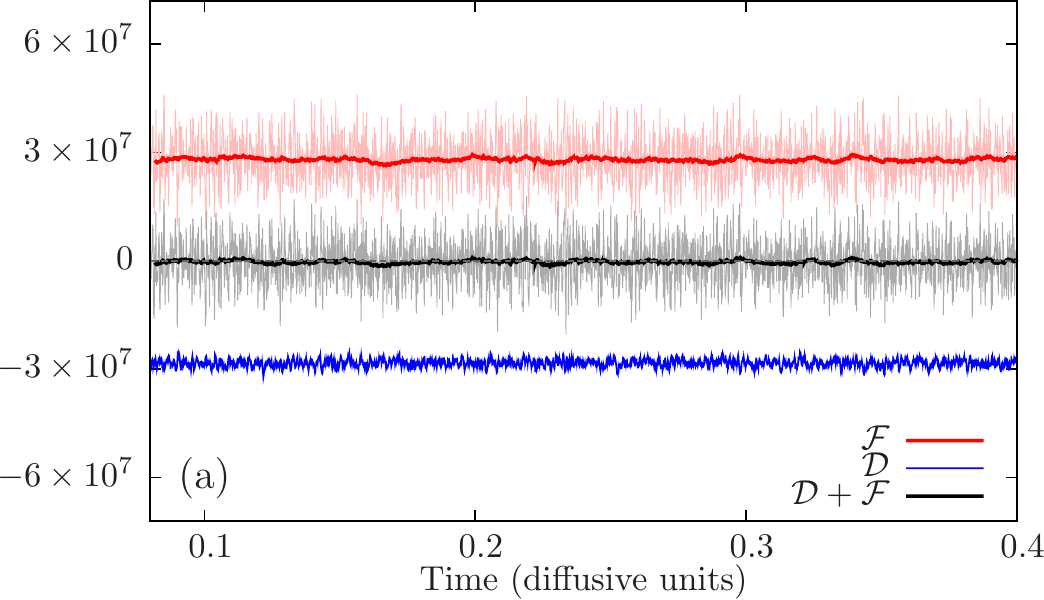}
   \hfill
   \includegraphics[width=0.49\textwidth]{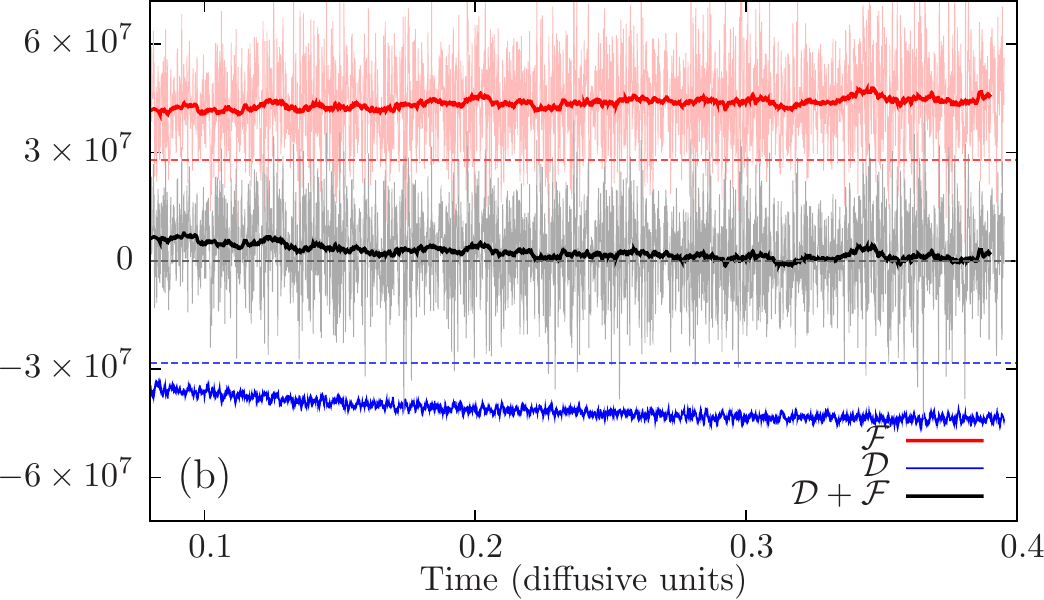}
   \caption{Time evolution of the forcing and dissipation terms for the 2D mode, as defined by Eq.~\eqref{eq:balance}. The instantaneous value is plotted as a thin transparent line while the thick opaque line corresponds to the time average over a window spanning $0.01$ vertical viscous time. (a) $A=0$. (b) $A=1600$. The horizontal dotted lines correspond to the time-averaged values in the reference case $A=0$.\label{fig:balance}}
\end{figure}

We examine next the energetics of the 2D depth-invariant flow.
Starting from Eqs.~\eqref{eq:momentum}-\eqref{eq:div}, the governing equations for the purely horizontal 2D flow $\left<\bm{u}\right>_z$ are (see also \cite{benavides_alexakis_2017})
\begin{align}
\label{eq:baro_mom}
\frac{\partial \left<\bm{u}\right>_z}{\partial t}+\left<\bm{u}\right>_z\cdot\nabla_h\left<\bm{u}\right>_z & =-Pr\nabla_h\left<p\right>_z+Pr\nabla_h^2\left<\bm{u}\right>_z-\left<\bm{u}'\cdot\nabla\bm{u}'\right>_z\,, 
\end{align}
where $\nabla_h$ is the horizontal gradient operator.
The last term in Eq.~\eqref{eq:baro_mom} corresponds to the forcing of the 2D mode by the 3D fluctuations.
Taking the scalar product of Eq.~\eqref{eq:baro_mom} with $\left<\bm{u}\right>_z$ and volume-averaging leads to the energy density equation of the 2D mode
\begin{equation}
\label{eq:balance}
\frac{d K_{2D}}{d t}=\underbrace{\frac{Pr}{\lambda^2} \iint \left<\bm{u}\right>_z\cdot\nabla_h^2\left<\bm{u}\right>_z \text{d}x \ \text{d}y}_{\displaystyle\mathcal{D}}+\underbrace{\left(-\frac{1}{\lambda^2}\iint \left<\bm{u}\right>_z\cdot\left<\bm{u}'\cdot\nabla\bm{u}'\right>_z \text{d}x \ \text{d}y\right)}_{\displaystyle\mathcal{F}} \ .
\end{equation}
The first term on the right-hand side corresponds to the viscous dissipation rate of the 2D flow while the last term represents the 2D energy production from the 3D fluctuations.
Looking at Eq.~\eqref{eq:balance}, it is clear that the 2D flow is in equilibrium only when the viscous dissipation $\mathcal{D}$ is balanced by the 3D forcing $\mathcal{F}$.
It follows that growth of $K_{2D}$ from a given equilibrium state can only be achieved by reducing the dissipation or increasing the forcing.
Figure~\ref{fig:balance}(a) shows that for $A=0$, the forcing is not zero but is exactly balanced by viscous dissipation.
The sum $\mathcal{D}+\mathcal{F}$ oscillates rapidly around zero, a fact consistent with the quasi-constant value of $K_{2D}$ observed for this case in figure~\ref{fig:kin1}(a).
For $A=1600$, however, both dissipation and forcing increase in amplitude, and the sum is on average positive at least initially, corresponding to the growth of the 2D kinetic energy observed in figure~\ref{fig:kin1}(a).
Note that the dissipation increases slowly with time while the baroclinic forcing term, while strongly fluctuating, remains quasi-constant (although it does grow very slightly).
This increase in the dissipation is only observed for the 2D barotropic component; the dissipation associated with the 3D fluctuations remains largely unchanged (not shown).
This is consistent with the condensation mechanism observed in purely 2D turbulence \citep{smith_yakhot1994,chertkov,Gallet2014} whenever no large scale damping term is introduced to balance the inverse energy flux, and is a consequence of the slow growth of the dominant energy-containing scale.  

\begin{figure}
   \vspace{2mm}
   \hspace{-5mm}
   \centering
   \includegraphics[width=0.74\textwidth]{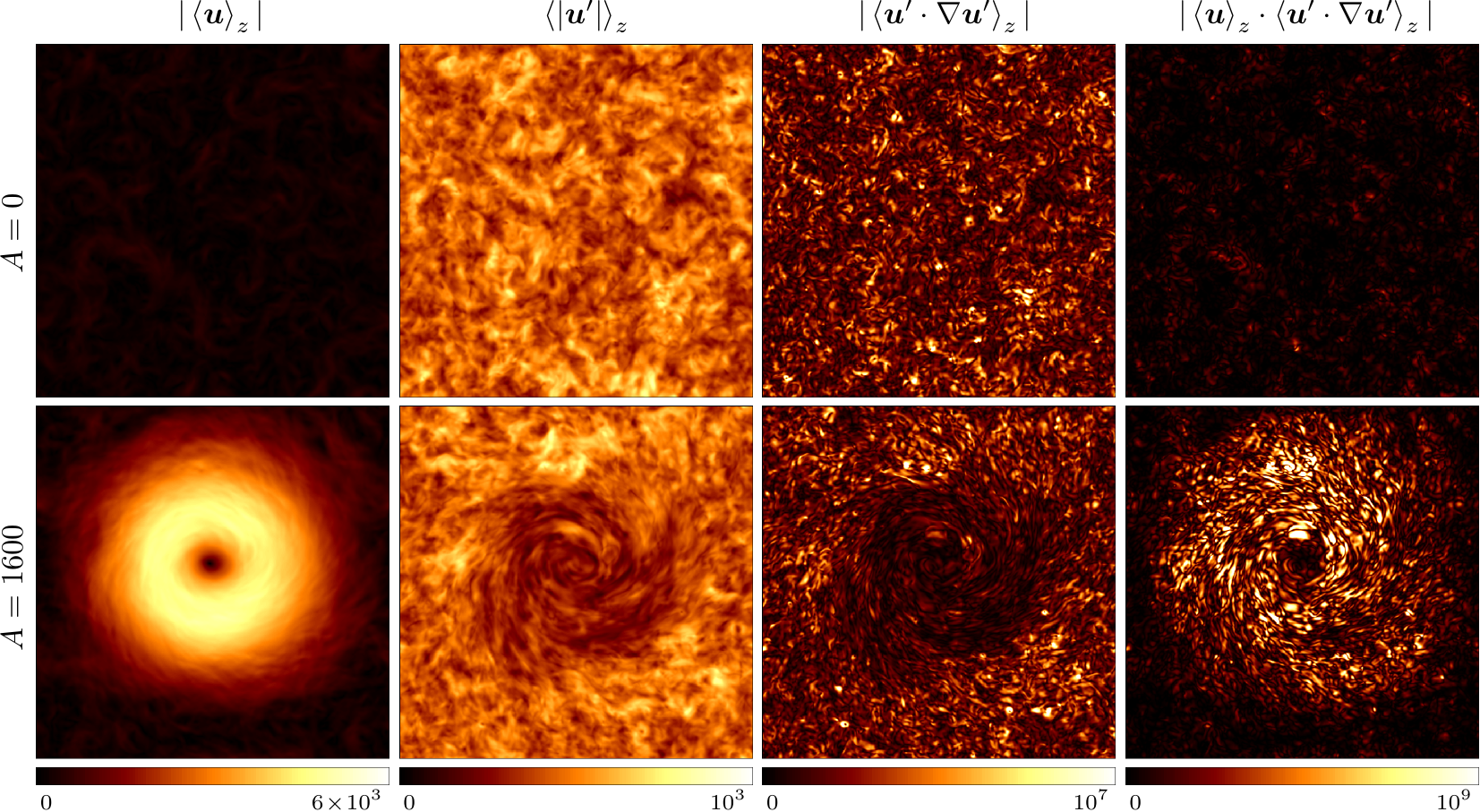}
   \hspace{1mm}
   \includegraphics[width=0.26\textwidth]{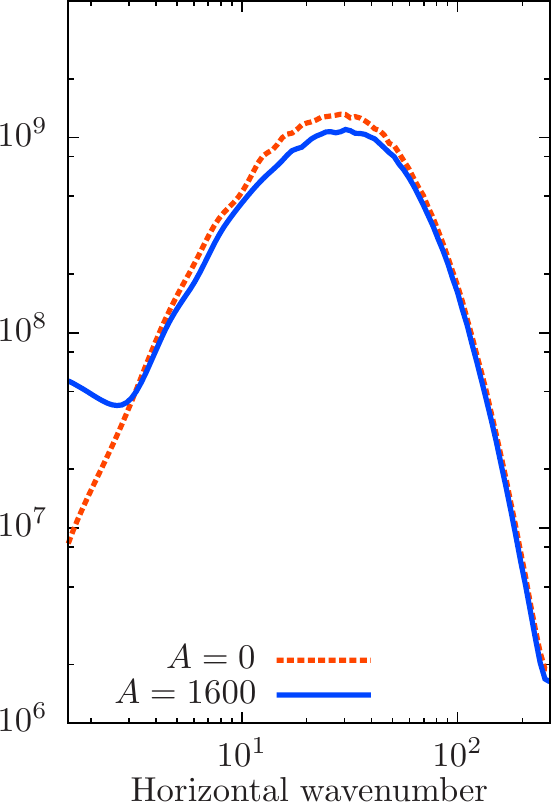}
   \caption{Left: amplitude of the 2D flow, vertical average of the 3D fluctuation amplitude, amplitude of the fluctuating forcing and its rate of working for $A=0$ (top line) and $A=1600$ (bottom line) at $t=0.3$. Right: power spectrum of the forcing $\left<\bm{u}'\cdot\nabla\bm{u}'\right>_z$ at the same time.\label{fig:forc}}
\end{figure}

At this stage, it is clear that the presence of an initially imposed vortex dipole somehow modifies the small scale convective flow, enhancing energy transfer into the 2D mode. One possible explanation for this behavior is that the 2D flow becomes significantly correlated with the 3D forcing, a correlation that can be quantified by looking at the angle $\gamma$ defined by $\cos\gamma=\left<\bm{u}\right>_z\cdot\left<\bm{u}'\cdot\nabla\bm{u}'\right>_z/(|\left<\bm{u}\right>_z| \ |\left<\bm{u}'\cdot\nabla\bm{u}'\right>_z|)$. However, our computations failed to reveal any significant differences in the probability density function of $\cos\gamma$ between cases with and without LSV (not shown), implying that the transfer enhancement cannot be explained by an increase in the correlation between the 3D forcing and the 2D flow.
Figure \ref{fig:forc} provides a clue.
The figure reveals a clear imprint of the large scale 2D structure $\left<\bm{u}\right>_z$ in the vertically-averaged 3D fluctuations $\left<|\bm{u}'|\right>_z$ and in the forcing term $\left<\bm{u}'\cdot\nabla\bm{u}'\right>_z$: the fluctuations are locally suppressed, likely by the strong, relatively ordered large scale shear or vorticity associated with the vortex structure, implying enhanced correlations in the phases of the small scale field.
This suppression of the small scale 3D fluctuations is consistent with the observed reduction in the Reynolds number (from $Re\approx535$ for $A=0$ to $Re\approx512$ for $A=1600$) and of the Nusselt number (from $Nu\approx28.8$ for $A=0$ to $Nu\approx27.6$), as already noted by \cite{guervilly_hughes_jones_2014} for control parameters for which the LSV emerges spontaneously.
It was also observed in thin-layer turbulence experiments \citep{Xia2011} although no subcritical behaviour was reported in this study.
This suppression in turn enhances interactions between two large $k_h$ 3D modes that transfer energy into a small $k_h$ 2D mode, bypassing the standard inverse cascade and enhancing the power spectrum of the forcing at these large scales, as shown in figure \ref{fig:forc} (right panel).
Note that, irrespective of the amplitude $A$, the forcing always peaks at $k_h\approx30$, which is approximately the Taylor microscale characteristic of the small scale vorticity field. The 2D energy is, however, very small at these scales (figure~\ref{fig:spect1}) and when $A=0$ this small scale forcing is in equilibrium with the dissipation (figure~\ref{fig:balance}(a)). For $A=1600$, however, the suppression of the small scale fluctuations leads directly to enhanced forcing of the 2D flow at the box scale, allowing for a runaway growth. The non-local nature of this upscale energy transfer has already been discussed in previous studies \citep{rubio2014,favier2014pof}, and we expect similar non-local energy fluxes in our subcritical state, although this remains to be fully explored in future studies.

The resulting positive feedback mechanism differs in an important respect from that present in 2D turbulence driven by a small scale {\it prescribed} stochastic force: in the present case the fluctuating force is itself dynamically modified by the LSV, and not just its rate of working as in \cite{Gallet2014}. In addition, the positive feedback revealed in figure~\ref{fig:forc} provides an indication that the LSV state will not in general spontaneously jump back to the lower LSV-free state: as soon as a transition starts to take place and the LSV is observed to grow, its back-reaction on the 3D fluctuations favours energy transfer into the 2D component, leading to run-away dynamics which can only be arrested by viscous condensation at the box scale or by other large scale effects not included in our simple model.

\vspace{-4mm}

\section{Discussion\label{sec:discuss}}

\begin{figure}
   \vspace{2mm}
   \centering
   \includegraphics[width=0.49\textwidth]{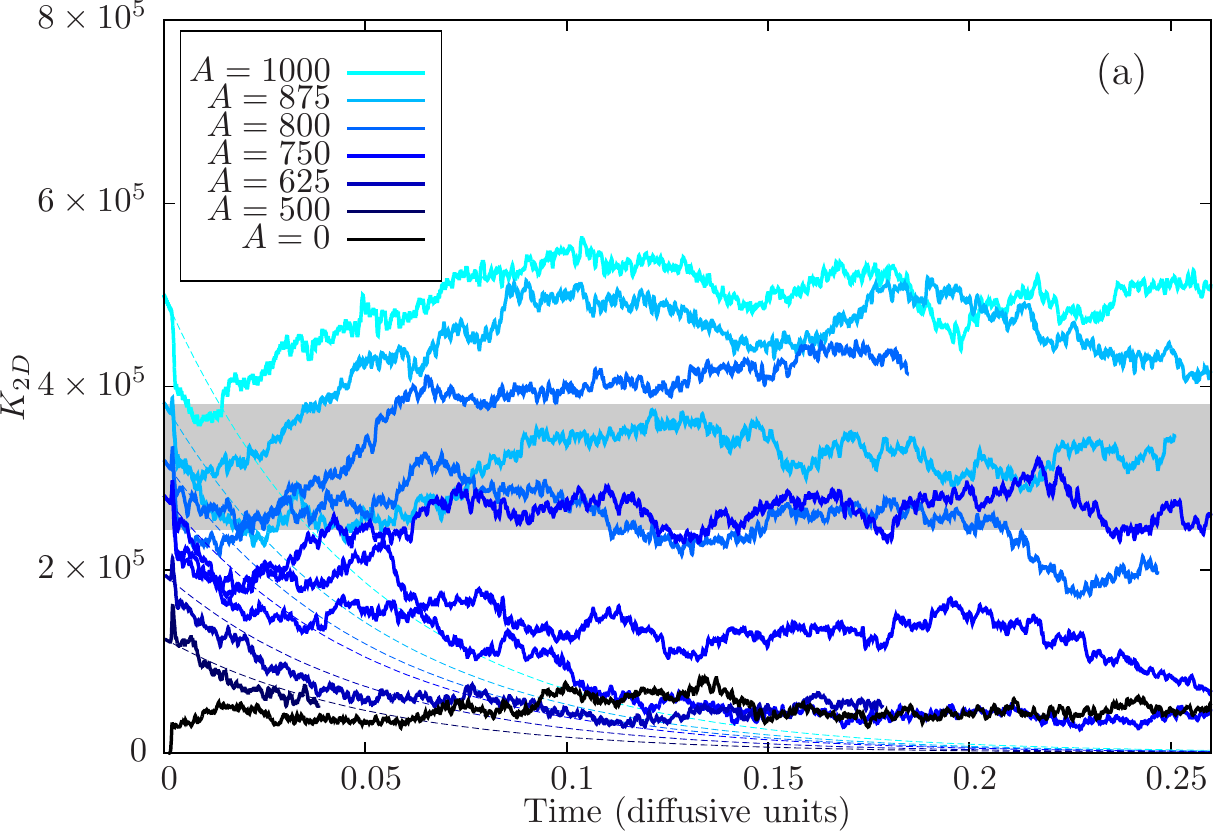}
   \hfill
   \includegraphics[width=0.49\textwidth]{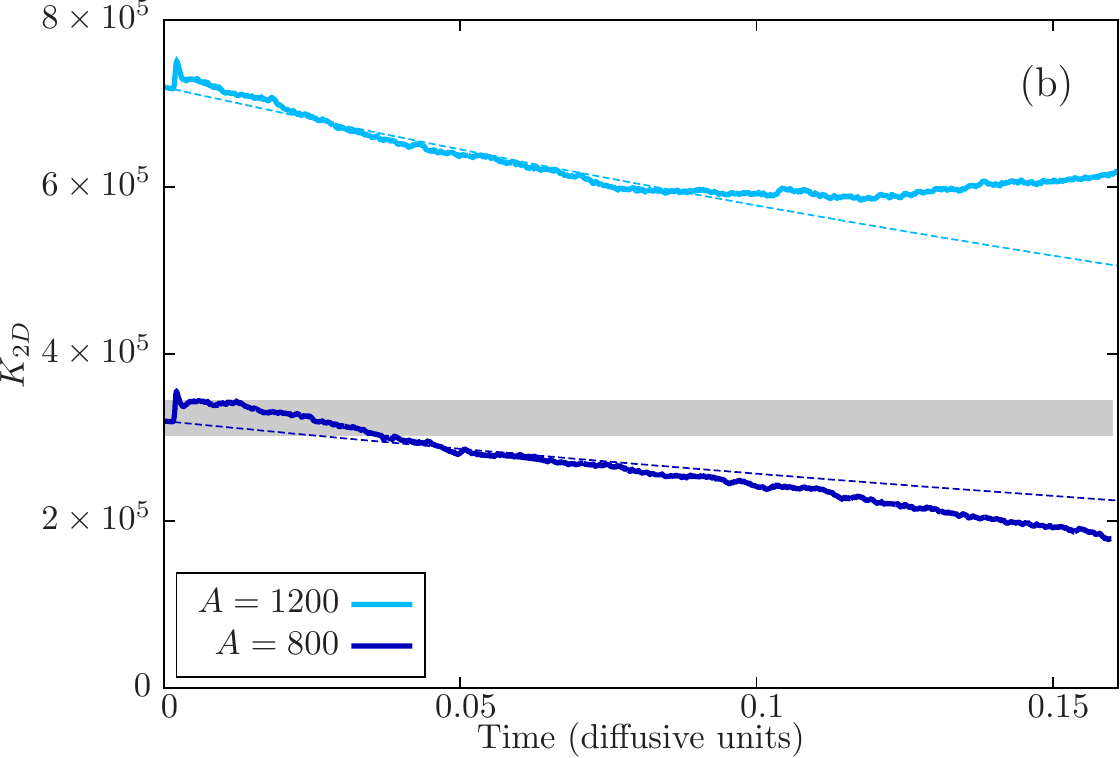}
   \caption{Time evolution of the kinetic energy density $K_{2D}$ of the 2D flow for different initial vortex amplitudes $A$ and aspect ratio $\lambda=2$ (a) and $\lambda=6$ (b). The shaded area corresponds to the values explored by $K_{3D}$.\label{fig:disc}}
\end{figure}

We now discuss how the subcritical behaviour identified here depends on the various control parameters. Thus far we have focused on aspect ratio $\lambda=4$, but similar simulations were performed with $\lambda=2$ (Table~\ref{tab:one}). We observe very similar results: the initial LSV is amplified only above a finite critical amplitude. For this value of the aspect ratio, however, the existence of a non-local inverse energy transfer is not systematic. When some of the simulations were repeated with different random initial temperature perturbations, some cases exhibited an inverse cascade while others did not (figure~\ref{fig:disc}(a)) in a manner reminiscent of pipe flow \citep{Darbyshire1995}. This is a consequence of the stochastic nature of the forcing term arising from the 3D fluctuations (figure~\ref{fig:balance}) which can drastically affect the properties of the transition \citep{Fauve2017}. Indeed the transition seems much less robust for $\lambda=2$ than for $\lambda=4$, an effect we ascribe to an increase in the amplitude of the fluctuations that arises from the smaller domain size, indicating that a reasonable scale separation between the initial vortex and the small scale flow is required for a robust and reproducible transition.
We confirmed this observation with simulations at $\lambda=6$, for which a case with $A=800$ decays while a case with $A=1200$ eventually grows (although with the available computational resources neither of these cases can be run until saturation owing to the presence of much longer transients, see figure~\ref{fig:disc}(b)).
This confirms that the observed transition is robust with respect to change in the aspect ratio of the vortex, although additional simulations (for example, decoupling the aspect ratio of the numerical domain and the initial size of the LSV) are required for a definitive conclusion about its effect on the threshold.

The results discussed above only apply for $Ta=10^8$ and $Ra=3\times10^7$.
We chose $Ta=10^8$ for numerical convenience, as increasing the Taylor number would be numerically prohibitive for the large aspect ratio domains and long-time dynamics considered here. Interestingly, the range of Rayleigh numbers for which a LSV is known to spontaneously emerge increases with the Taylor number \citep{guervilly_hughes_jones_2014,favier2014pof}. It is reasonable, therefore, to assume that subcritical transitions will also be observed at higher Taylor numbers, and might even be more prominent. However, this hypothesis remains to be confirmed.

We have also explored the effect of varying the Rayleigh number. First, it is known that a LSV cannot be sustained when the convection is insufficiently turbulent. For $Ta=10^8$, $Ra\ge5\times10^6$ is required: when we repeated our simulations for $Ra=4\times10^6$, varying the amplitude $A$ of the vortex, we found no subcritical behaviour and all simulations eventually converged towards the same equilibrium dominated by 3D fluctuations. It appears that the subcritical transition is only present for sufficiently turbulent flows, well above the threshold of the linear instability. Finally, the same experiments were repeated for higher Rayleigh numbers (namely $Ra=4\times10^7$ and $Ra=5\times10^7$) with results that are more subtle: for $\lambda=2$, we did not observe any subcriticality, obtaining a 3D state irrespective of $A$. Surprisingly, for $\lambda=4$ and $Ra=4\times10^7$, we recovered a LSV state that was not present for $\lambda=2$. This fact provides further evidence that a clear scale separation between the vortex and the small scale 3D flow favors a subcritical transition. A detailed study of the scaling of the critical amplitude $A_c$ with $Ta$, $Ra$ and $\lambda$ is, however, beyond the scope of this paper, although it represents an interesting line of investigation not only from a fundamental point of view, but also for possible applications to LSV structures in geophysical and astrophysical flows.

Our choice of initial conditions \eqref{eq:init} is arbitrary and other options could be explored. In our moderate Rossby number simulations, symmetry breaking favouring cyclonic motions has been observed, in contrast with the symmetry between cyclonic and anticyclonic vortices found in the limit $Ro\to 0$ \citep{julien2012}. Our initial condition is, however, perfectly symmetric requiring a significant transient phase to break up the large but unstable anticyclone, something that could be avoided if the calculations were to be initialized with a cyclonic structure only. In this respect, the reduced description of \cite{julien2012} may be useful for exploring the detailed mechanism behind the subcritical transition discovered in this paper and in particular the presence of hysteresis in $Ra$, provided, of course, that the reduced equations exhibit similar behaviour. 

We believe that the mechanism responsible for the observed subcritical transition is not specific to rotating convection and is likely to occur in other systems with multiple cascade scenarios, including thin-layer and magnetohydrodynamic turbulence.
In particular, we emphasize that rotation is not required to observe LSV in 3D flows.
Small scale anisotropy, and its possible enhancement by large scale flows, is the key and is present in all turbulent systems with multiple cascade scenarios, from thin-layer to magnetohydrodynamic turbulence.
We also emphasize that the transition identified here separates two fully turbulent states, in contrast to the classical subcritical transition from laminar to turbulent shear flow in, e.g., pipe flow \citep{Darbyshire1995,eckhardt2007}. Some concepts developed in this field may nevertheless prove useful in the present context, such as the search for optimal perturbations \citep{kerswell2018}.

\vspace{-2mm}
\section*{Acknowledgments} This work was initiated during the workshop ``Rotating Convection: from the Lab to the Stars'' organized and supported by the Lorentz Center at the University of Leiden (\url{http://www.lorentzcenter.nl/}).
This work was granted access to the HPC resources of Aix-Marseille Universit\'e financed by the project Equip@Meso (ANR-10-EQPX-29-01) of the program ``Investissements d'Avenir'' supervised by the Agence Nationale de la Recherche.
Computations were also conducted with the support of the HPC resources of GENCI-IDRIS (Grants No. A0020407543 and A0040407543), the Rocket High Performance Computing service at Newcastle University and the ARCHER UK National Supercomputing Service (\url{http://www.archer.ac.uk}).
C.G. was supported by the UK Natural Environment Research Council under grant NE/M017893/1.
B.F. acknowledges funding by the European Research Council under the European Union's Horizon 2020 research and innovation program through Grant No. 681835-FLUDYCO-ERC-2015-CoG.

\bibliography{biblio}
\bibliographystyle{jfm}

\end{document}